\documentclass[useAMS,usenatbib]{mn2e}
\usepackage{amssymb}
\usepackage{epsfig}
\usepackage{times}
\usepackage{longtable} 

\title[Strong size evolution of the most massive galaxies since z$\sim$2]
{Strong size evolution of the most massive galaxies since z$\sim$2}

\author[I. Trujillo et al.]{Ignacio Trujillo$^{1,2}$, Christopher J. Conselice$^{1}$,
 Kevin Bundy$^{3}$, M. C. Cooper$^{4}$,\newauthor  P. Eisenhardt$^{5}$  and Richard S.  Ellis$^{6}$
\\
$^{1}$ School of Physics and Astronomy, University of Nottingham, University Park, Nottingham NG7 2RD, UK\\
$^{2}$ Present address: Instituto de Astrof\'isica de Canarias, V\'ia L\'actea s/n, 38200 La Laguna, Tenerife, Spain\\	
$^{3}$ Dept. of Astronomy and Astrophysics, University of Toronto, 50 St. George
Street, Rm 101, Toronto, ONM5S 3H4, Canada\\
$^{4}$Department of Astronomy, University of California at Berkeley, Berkeley, CA 94720, USA \\
$^{5}$Jet Propulsion Laboratory, California Institute of Technology, Pasadena, CA 91109, USA \\
$^{6}$Caltech MC 105-24, 1201 East California Boulevard, Pasadena, CA 91125, USA}

\begin{document}

\date{}

\pagerange{\pageref{firstpage}--\pageref{lastpage}} \pubyear{2002}

\maketitle

\label{firstpage}

\begin{abstract}

Using the combined capabilities of the large near-infrared Palomar/DEEP-2 survey, and the superb
resolution of the ACS HST camera, we explore the size evolution of 831 very massive galaxies
(M$_\star$$\ge$10$^{11}$h$_{70}$$^{-2}$M$_{\sun}$) since z$\sim$2. We split our sample according 
to their light concentration using the S\'ersic index $n$. At a given stellar mass, both low
(n$<$2.5) and high (n$>$2.5)  concentrated objects were much smaller in the past than their local
massive counterparts. This evolution is particularly strong for the highly concentrated
(spheroid-like) objects.  At z$\sim$1.5, massive spheroid-like objects were a factor of 4($\pm$0.4)
smaller (i.e. almost two orders of magnitudes denser) than those we see today. These small sized,
high mass galaxies do not exist in the nearby Universe, suggesting that this  population merged
with other galaxies over several billion years to form the largest galaxies we see today.

\end{abstract}


\section{Introduction}

In the nearby Universe, the population of galaxies with stellar masses greater than 10$^{11}$
M$_{\sun}$ is dominated by large early--type galaxies (Baldry et al. 2004) with correspondingly
large sizes (Shen et al. 2003). These nearby systems contain old and metal-rich stellar populations
that formed quickly in the early Universe (Heavens et al. 2004; Feulner et al. 2005; Thomas et al.
2005). However, it has remained unknown whether the stars in these galaxies were all assembled in
the same system, or formed in lower mass galaxies that later merged.  

Two major formation models have been proposed in order to explain the properties of these
galaxies: the so-called monolithic collapse model  (Eggen, Lynden-Bell \& Sandage 1962; Larson
1975; Arimoto \& Yoshii 1987; Bressan, Chiosi \& Fagotto 1994) and the hierarchical merging
scenario (Toomre 1977; White \& Frenk 1991). These two mechanisms have both observational and
theoretical successes and drawbacks. For instance, evidence in favor of a fast and dissipative
monolithic collapse is the fact that the bulk of stars in massive ellipticals are old (Mannucci
et al. 2001) and have high [$\alpha$/Fe] ratios (i.e. short star formation time-scales; Worthey
et al. 1992). In addition, the structural and dynamical properties of observed spheroid galaxies
are well reproduced by cold dissipationless collapse (van Albada 1982; May \& van Albada 1984;
McGlynn 1984; Aguilar \& Merritt 1990; Londrillo et al. 1991; Udry 1993; Hozumi et al. 2000;
Trenti et al. 2005), a process that is expected to dominate the last stages of a highly
dissipative collapse. On the other hand, supporting a hierarchical merger scenario, observations
find a decline in the number of massive galaxies seen at higher redshifts. This decline is moderate since
z$\sim$1 and much stronger at even higher redshifts (Daddi, Cimatti \& Renzini 2000; Pozzetti et al. 2003;
Bell et al. 2004; Drory et al. 2004, 2005; Daddi et al. 2005; Saracco et al. 2005; Faber et al.
2005; Pozzetti et al. 2007). Moreover, new renditions of semi-analytical models where
merging is the cornerstone of galaxy formation (e.g. De Lucia et
al. 2006) are much better able to match the stellar population properties of elliptical galaxies.
Finally, it has been found theoretically that the scaling laws followed by elliptical galaxies
are robust against merging (Ciotti et al. 2007).

Exploring the assembly of massive galaxies with cosmic time through number density analysis or
merger rate estimations are however difficult to conduct since, first, very massive galaxies are scarce
and, second, their clustering properties make them strongly affected by field-to-field variance
associated with limited observed volumes. These two difficulties make claims of the assembly of
massive galaxies in the early Universe based on number densities and merger rate estimations
highly uncertain (e.g., Cimatti, Daddi \& Renzini 2006; Renzini 2007).

A much more straightforward approach to the assembly problem of massive galaxies is to explore
the size evolution of these systems at a given stellar mass. In a "monolithic-like" scenario 
where a galaxy is fully assembled after the formation of its stars, the stellar mass-size
relation should remain unchanged as cosmic time evolves. In the hierarchical merging scenario,
however, the stellar mass-size relation will evolve as a result of the increase in size after
each galaxy merger. For instance, state-of-the-art hierarchical semi-analytical models predict a
very strong (a factor of 1.5-3) evolution in the size of very massive galaxies
(M$_\star$$\ge$10$^{11}$h$_{70}$$^{-2}$M$_{\sun}$) nine billion years ago (Khochfar \& Silk
2006b). This predicted evolution in size is a strong function of galaxy mass, with the more
massive objects expected to have the largest increase in size.

Due to the lack of a large sample of very massive galaxies at high redshift, analysis of the
evolution of the stellar mass-size relation of galaxies have mainly explored objects in the
10$^{10}$$<$M$_{\star}$$<$10$^{11}$ M$_{\sun}$ range. Observations within this mass interval do not
find a significant evolution of the stellar mass-size relation since z$\sim$1 (Barden et al. 2005;
McIntosh et al. 2005). Explorations of these objects at even earlier look-back times (1$<$z$<$3)
have been also attempted, and show a moderate decline in size for galaxies at a given
stellar mass (Trujillo et al. 2004; 2006a).  However, the size of very massive galaxies,
M$_\star$$\ge$10$^{11}$M$_{\sun}$, is still largely unexplored, with only around a dozen objects
studied in detail so far (Daddi et al. 2005; Trujillo et al. 2006b; Longhetti et al. 2007; McGrath
et al. 2007). Definitive conclusions regarding the evolution of sizes for massive galaxies thus
remain largely unknown. The goal of this paper is to shed some light on this issue by exploring the
sizes of a large sample of  very massive galaxies since z$<$2. This will help to clarify whether
the evolutionary scenario for these objects is hierarchical or monolithic-like.

The paper is structured as follows. In Section 2 we give a brief summary of the Palomar/DEEP2 data,
and in Section 3 we describe the determination of the stellar masses. Size measurement technique
and robustness estimations for our data are provided in Section 4. In Section 5 we study the
selection effects and in Section 6 present the observed stellar mass--size relations. We compare
our results with other samples in Section 7, and finally, we discuss our results in Section 8.  In
what follows, we adopt a cosmology of  $\Omega_m$=0.3, $\Omega_\Lambda$=0.7 and H$_0$=70 km
s$^{-1}$ Mpc$^{-1}$.

\section[]{Description of the data}

We use the Palomar Observatory Wide-field Infrared (POWIR)/DEEP2 survey 
(Bundy et al. 2006; Conselice et al. 2007a,b; Davis et al. 2003) 
to define a sample of 831 galaxies with masses larger than 10$^{11}$h$_{70}$$^{-2}$M$_{\sun}$ located 
over $\sim$710 arcmin$^2$ in the Extended Groth Strip (EGS). This field (63 Hubble Space Telescope tiles) was
imaged with the Advanced Camera for Surveys (ACS) in the V(F606W, 2660s) and I-band (F814W, 2100s). Each
tile was observed in 4 exposures that were combined to produce a pixel scale of 0.03 arcsec, with a
Point Spread Function (PSF) of 0.12 arcsec Full Width Half Maximum (FWHM).  In addition to the HST
data, optical imaging from the CFHT 3.6-m telescope in the B, R and I bands taken with the CFH12K
camera was used. Integration times for these observations were 1 hour in B and R, and 2 hours in I. 
Limiting magnitudes reached are B=24.5 (AB, 8$\sigma$), R=24.2 (AB, 8$\sigma$) and I=23.5 (AB,
8$\sigma$). The details of the data reduction for this data is provided in Coil et al. (2004).

  In the EGS region, the Palomar Near-Infrared K$_s$-band imaging has a typical depth of greater
than K$_{AB}$=22.5 mag (5 $\sigma$), and a J-band depth of J$_{AB}$=23.4 mag (5 $\sigma$). About
70\% of the galaxies brighter than R$_{AB}$=24.1 mag in this field were targeted by the DEEP2
Galaxy Redshift survey, using the Keck 10m telescope, for a total spectroscopic redshift
completeness of $\sim$56\%. In our case, 410 objects have a spectroscopic redshift ($\sim$50\% of
the full dataset), with a spectroscopic completeness of 61\% for galaxies with z$<$1, and a
completeness of 30\% for galaxies with z$>$1. We supplemented our spectroscopic redshift catalogue
with photometric redshifts with an accuracy for all galaxies in the total K-band survey  of
$\delta$z/(1+z)$\approx$0.07 in the range 0.2$<$z$<$1.4 and $\delta$z/(1+z)$<$0.22 for galaxies at
1.4$<$z$<$2 (Conselice et al. 2007a).  However, it is worth noting that, for the massive galaxies
we use in this paper, the accuracy is $\delta z/(1+z) = 0.025$ at $z < 1.4$, and likely similarly
lower for $z > 1.4$ galaxies (i.e. $\delta z/(1+z) \approx  0.08$). 

\section[]{Determination of stellar masses}

The determination of stellar masses for each of our galaxies follows a standard multi-color (BRIJK)
stellar population fitting technique, producing uncertainties of $\approx$0.2 dex. The largest
systematic source of error comes from the assumed Initial Mass Function (IMF), in this paper we
have used the Chabrier IMF for all stellar mass measurements. The details of these estimations are
presented in Bundy et al. (2006) and Conselice et al. (2007b).  However, because the measurement of
stellar masses is an integral part of this paper, we give a brief description of how our masses are
measured, and what systematics might be present.

The basic mass determination method we use consists of fitting a grid of model SEDs 
constructed from Bruzual \& Charlot (2003) (BC03) stellar population synthesis models, 
with different star formation histories.  We use an exponentially declining model
to characterise the star formation, with various ages, 
metallicities and dust contents included.  These models are parameterised
by an age, and an e-folding time.  These parameterisations are
fairly simple, and it remains possible that stellar mass from
older stars is missed under brighter younger populations. While
the majority of our systems are passively evolving older stellar populations,
it is possible that up to a factor of two in stellar mass is missed in any 
star bursting blue systems.  However, stellar masses measured through
our technique are roughly the expected factor of 5-10 smaller than
dynamical masses at $z \sim 1$ using a sample of disk galaxies
(Conselice et al. 2005), demonstrating their inherent reliability.

We calculate 
the likely stellar mass, age, and absolute magnitudes for each galaxy at all 
star formation histories, and determine stellar masses based on this
distribution.  Distributions with larger ranges of stellar masses
have larger resulting uncertainties. It turns out that while parameters such 
as the age, e-folding time, metallicity, etc. are not likely accurately fit
through these
calculations due to various degeneracies, the stellar mass is robust.  
Typical errors for our stellar masses are 0.2 dex from the width
of the probability distributions.  There are also uncertainties from
the choice of the IMF.  Our stellar masses utilise the
Chabrier IMF, which can be converted to Salpeter IMF stellar masses 
by adding 0.25 dex.  There are additional
random uncertainties due to photometric errors.  The resulting
stellar masses thus have a total random error of 0.2-0.3 dex,
roughly a factor of two. 

There is furthermore the issue of whether or not our stellar masses are overestimated based on
using the Bruzual \& Charlot (2003) models.  It has recently been argued by Maraston (2005) and
Bruzual (2007) that the exclusion of an updated treatment of thermal-pulsating AGB stars  
in the BC03 models results in
calculated stellar masses  too high by a factor of a few. While we consider an uncertainty of a
factor of two in our stellar masses, we must consider whether our sample is in the regime where the
effects of TP-AGB stars will influence our mass measurements.  This has been investigated recently
in Maraston (2005) and Bruzual (2007) who have both concluded that galaxies stellar masses computed 
using newer TP-AGB star prescriptions are up to roughly 50-60\% lower than without. This is
particularly true for masses determined in the rest-frame infrared.   

This problem has also been recently investigated independently by Kannappan \& Gawiser (2007) who
come to similar conclusions, but do not advocate one model over another.  Furthermore, the effect
of TP-AGB stars is most pronounced in the rest-frame IR, and for young stellar population ages. Our
survey is K-selected, and the observed K-band is used as the flux in which the masses are computed.
The rest-frame wavelength probed with the observed K-band ranges from 0.7$\mu$m to 1.5$\mu$m where
the effects of TP-AGB stars are minimised. The ages of our galaxies are also older than the ages
where TP-AGB stars have their most effect (Maraston 2005; Bruzual 2007). It is also worth noting
that the effects of TP-AGB stars are more important when normalising stellar masses further into
the red.  In this paper we find a strongest evolution at higher redshifts where we are probing the
rest-frame optical at observed-K, and where the effects of TP-AGB stars are minimised (e.g.,
Bruzual 2007).   We investigate  the effects of TP-AGB stars in our estimations by determining how
our stellar masses change between using the BC03 models and the updated Charlot \& Bruzual (2007,
in prep) models using the new TP-AGB methods. We find for 1330 massive galaxies in the Palomar
sample a difference of $\sim$0.07 in log M$_{*}$, which is neglectable. A similar conclusion has
been recently achieved by McGrath, Stockton \& Canalizo (2007) using a sample of galaxies at
z$\sim$1.5 with stellar masses similar to those explored here. On comparing the new rendition of
models from Charlot \& Bruzual (2007) vs BC03, they found that the inferred galaxy masses are
slightly ($\sim$10\%) smaller using the 2007 models.

Although we do have Spitzer data for our sources, we do not
use this imaging for two reasons. The first is that by normalising
the stellar masses with  Spitzer IRAC magnitudes, we are
in a regime where the TP-AGN stars are more pronounced, and
thus would affect the stellar mass measurements to a degree
even greater than using the observed K-band (Kannappan \&
Gawiser 2007; Bruzual 2007).  Secondly, the large PSF of the
IRAC images makes it difficult to obtain accurate photometry
for many of our sources due to contamination from other
galaxies with overlapping PSFs.  While it is possible to correct
for this, the large resulting random photometry uncertainties 
make the stellar masses less certain.

Another possible source of uncertainty is the photometric redshifts we use for our sample. While at
$z < 1.4$ about half of our sample have spectroscopic redshifts, at $z > 1.4$ all of our systems
have photometric redshifts. We can however determine the accuracy for those systems at $z < 1.4$. 
The agreement is very good for our massive systems with $\delta z/(1+z) = 0.025$. This results in
another $<$20\% uncertainty in the stellar mass measurements. For $z > 1.4$, if we assume $\delta
z/(1+z) \approx 0.08$, the uncertainty in the stellar masses would be $\sim$32\%. Overall, however,
these uncertainties cannot account for the trends seen later in this paper.

\section[]{Size estimation}

The structural parameters used in this paper were measured using the ACS I-band filter.  Sizes
(as parameterised by the half-light or effective radius along the semi-major axis a$_e$) were
estimated using the GALFIT code (Peng et al. 2002). Sizes were circularised,
r$_e$=a$_e$$\sqrt{(1-\epsilon)}$ , with $\epsilon$ being the ellipticity of the object. 

GALFIT convolves S\'ersic (1968) r$^{1/n}$ galaxy models with the PSF of the images, and
determines the best fit by comparing the convolved model with the galaxy surface brightness
distribution using a Levenberg-Marquardt algorithm to minimise the $\chi^2$ of the fit. The
S\'ersic model is a flexible parametric description of the surface brightness distribution of
the galaxies and contains the exponential (n=1) and de Vaucouleurs (n=4) models as particular
cases. In addition, this model is used in the structural analysis of the SDSS galaxy sample
(our local comparison sample; Blanton et al. 2003; Shen et al. 2003).

The S\'ersic index n measures the shape of surface brightness profiles. In the nearby
Universe, galaxies with n$<$2.5 are mostly disk-like objects, whereas galaxies with n$>$2.5 are
mainly spheroids (Ravindranath et al. 2002). We use this S\'ersic index criterion to split our
sample at higher redshifts and facilitate a comparison with the local galaxy population. During
the fit, neighbouring galaxies were excluded using a mask, but in the case of closely
neighbouring objects with overlapping isophotes, the objects were fitted simultaneously. The
results of our fitting are shown in the Appendix: Table \ref{rawdata}.

\subsection{Testing the structural parameters estimates: simulations}

\begin{figure}
\epsfig{file=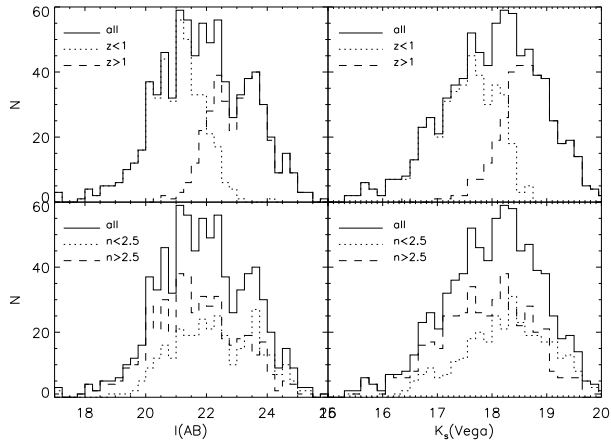,width=0.45\textwidth} 
\centering

\caption{Histograms showing the magnitude distributions in I(AB) and K$_s$(Vega) bands of the
galaxies in our sample. The sample is split by redshift and S\'ersic index.}

\label{maghist}
\end{figure}

The results presented in this paper rely on our ability to measure accurate structural parameters.
To gauge the accuracy of our parameter determination realistic simulations were conducted. We have
created 1000 artificial galaxies uniformly generated at random in the following ranges, matching
the observed distribution of our galaxies: 18$\leq$I$_{AB}$$\leq$26 (see Fig. \ref{maghist}),
0.03$\leq$r$_e$$\leq$2.55 arcsec (which at z$\sim$1 equals 0.25-20.5 h$_{70}^{-1}$ kpc),
0.5$\leq$n$\leq$8, and  0$\leq$$\epsilon$$\leq$0.8. To simulate the real conditions of our
observations, we add a background sky image taken (randomly at each time) from a piece of the ACS
I-band image. Finally, the galaxy models were convolved with the observed PSF. The same procedure
was used to retrieve the structural parameters in both the simulated and actual images. 

\begin{figure*}
\epsfig{file=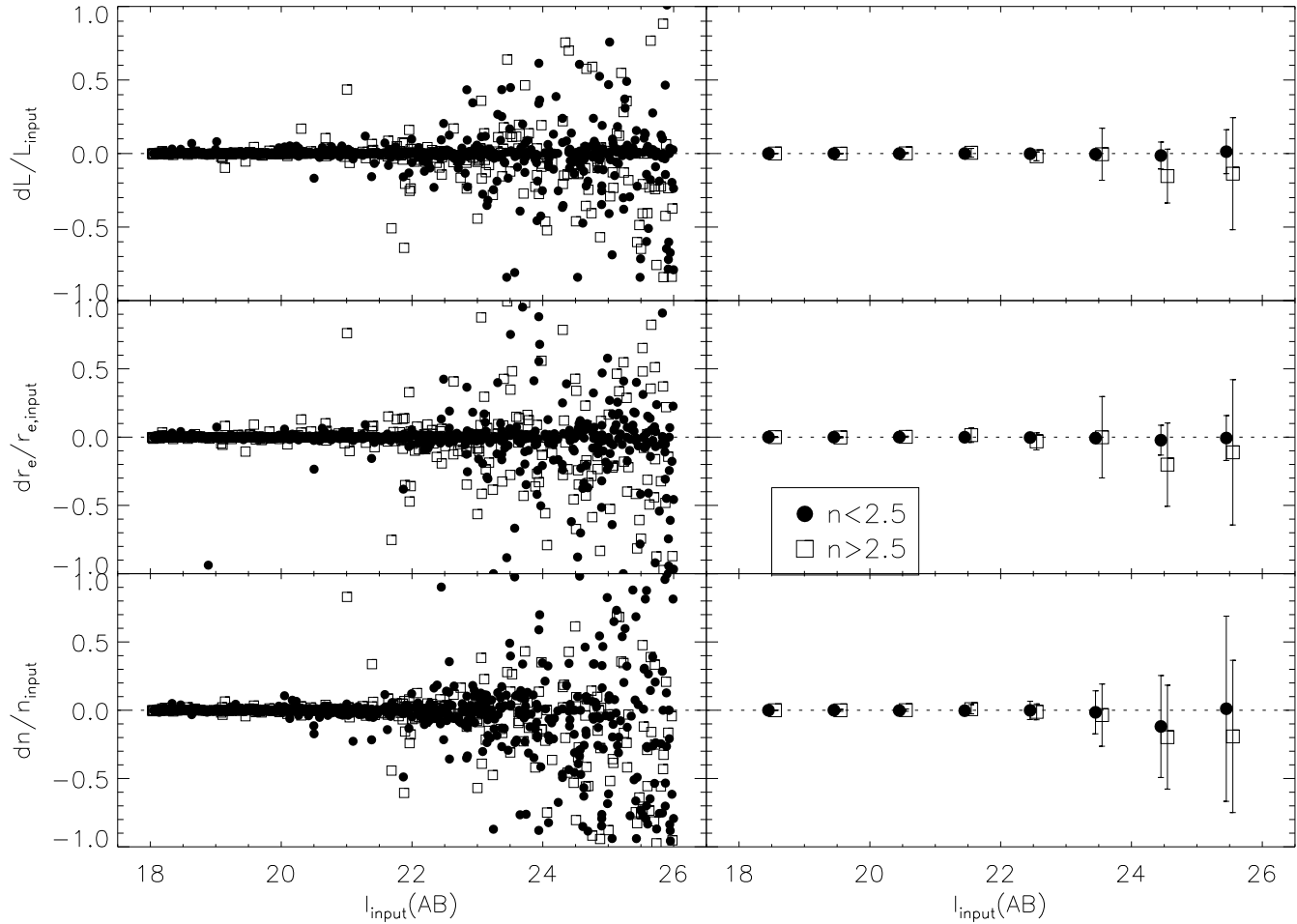,width=\textwidth} 

\centering

\caption{Relative errors derived from the difference between the input and recovered structural
parameters [(output-input)/input] according to our simulations on the HST F814W (I-band) 
imaging. Filled
symbols are used to indicate less concentrated objects (n$_{\rm input}$$<$2.5), whereas open symbols
imply highly concentrated objects (n$_{\rm input}$$>$2.5). The right panels show the mean systematic
difference and 1$\sigma$ error bars.}

\label{simul1}
\end{figure*}

\begin{figure*}
\epsfig{file=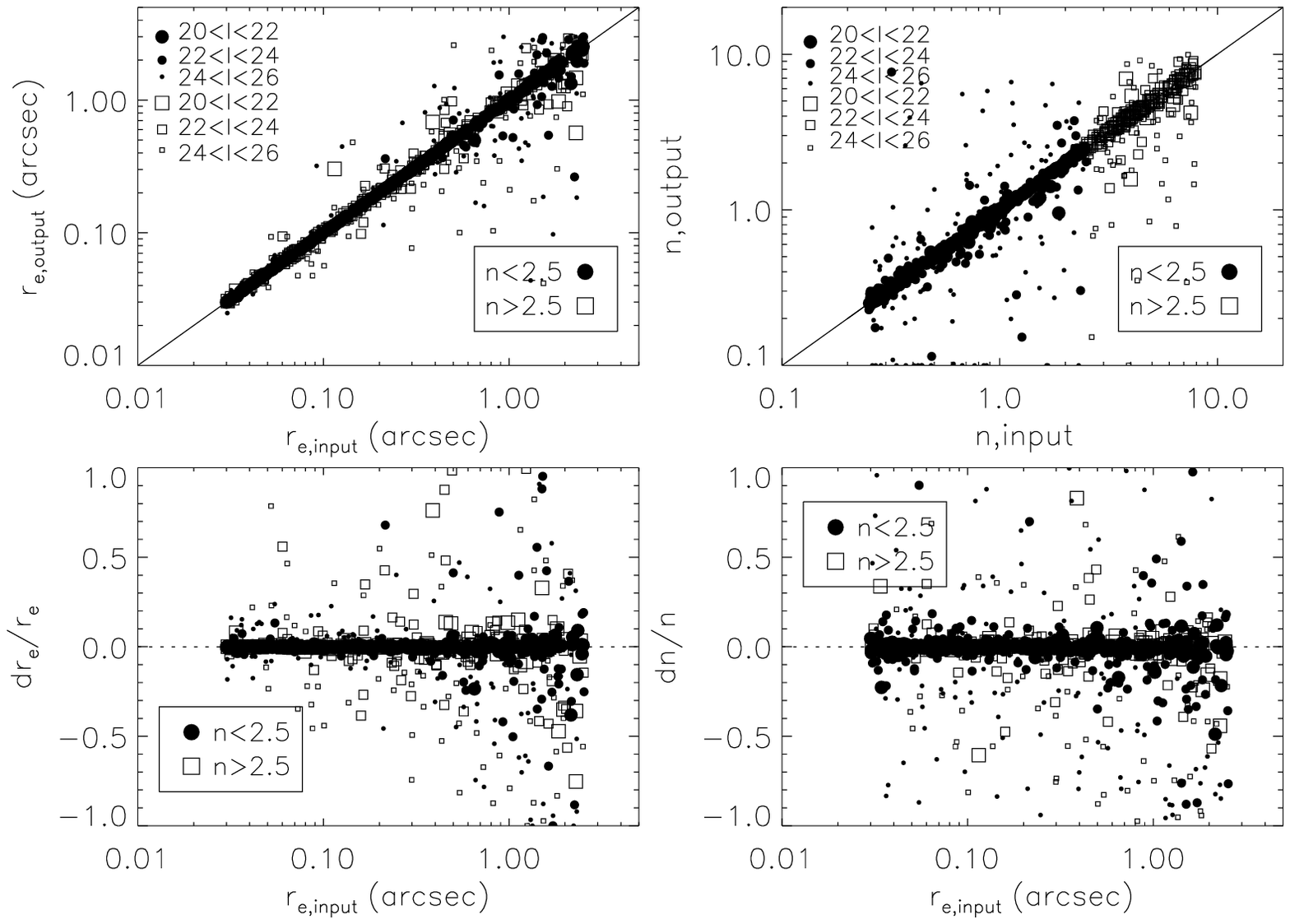,width=\textwidth} 
\centering

\caption{Galaxy size-measurement bias: the figure shows a comparison between input and recovered
structural parameter values in our simulations with the F814W (I-band) (AB) imaging. Top left: Relation
between measured and the input intrinsic half-light radius (before seeing convolution). Top right:
Relation between measured and input seeing deconvolved S\'ersic index n. Bottom left: Relative
error between the input and the measured seeing deconvolved effective radius
[dr$_e$/r$_e$=(r$_{e,output}$-r$_{e,input}$)/r$_{e,input}$] vs. the input effective radius. Bottom
Right: Relative error between the input and the measured seeing deconvolved S\'ersic index n
[dn/n=(n$_{output}$-n$_{input}$)/n$_{input}$] vs. the input effective radius. Filled symbols are
used to indicate less concentrated objects (n$_{input}$$<$2.5), whereas open symbols imply highly
concentrated objects (n$_{input}$$>$2.5).}

\label{simul2}
\end{figure*}

The results of these simulations are shown in Figures \ref{simul1} and \ref{simul2}. Fifty percent
of our galaxies are brighter than I(AB)=21.9 mag, and for these galaxies we find dr$_e$/r$_e$$<$5\%
and dn/n$<$7\%. For 95\% of our galaxies which are at I(AB)$<$24.2 mag, the uncertainties are
dr$_e$/r$_e$$<$30\% and dn/n$<$38\%. As expected, at fainter apparent magnitudes the structural
parameters are recovered with larger uncertainties. Only for magnitudes fainter than I(AB)=24 mag
is there a small bias ($\sim$20\%) of the index n towards smaller indices for galaxies with
n$_{input}$$>$2.5. Recovery of the structural parameters with larger n are more affected than
those with lower values. As shown in Figure \ref{simul2}, we do not find any bias of our sizes or
shape parameter as a function of the size of the objects. Finally, we have also explored whether
the variation of the PSF along the image can affect the recovery of the size of the galaxies. Using
different stars in the images as a PSF we find that the estimation of the sizes is robust to
changes in the selected PSF to analyze the data: the scatter between the sizes is $\lesssim$10\%
(1$\sigma$).

\subsection{Potential sources of systematic errors: K-correction effects and AGN contamination}

The sizes of the galaxies presented in this paper were measured using the ACS observed I-band,
which implies that for galaxies at z$>$1.3 sizes are retrieved in the rest-frame UV. To check
whether this K-correction effect can affect our size estimates we have compared our sizes measured
in the I-band with sizes obtained in the Near Infrared Camera and Multi-Object Spectrometer
(NICMOS) H-band (F160W) for a subset of 27 galaxies (with 0.8$<$z$<$1.8 and median z$\sim$1.2)
that were observed at both wavelengths. These NICMOS data consist of 63 pointings of camera 3
(52$\times$52 arcsec, 0.203 arcsec/pixel) in the EGS field. Each pointing is the combination of
four sub-pixel dithered exposures, with a total exposure time of 2600 sec. The final mosaic was
assembled using a drizzle task and has a pixel scale of 0.10 arcsec. 

\begin{figure*}
\epsfig{file=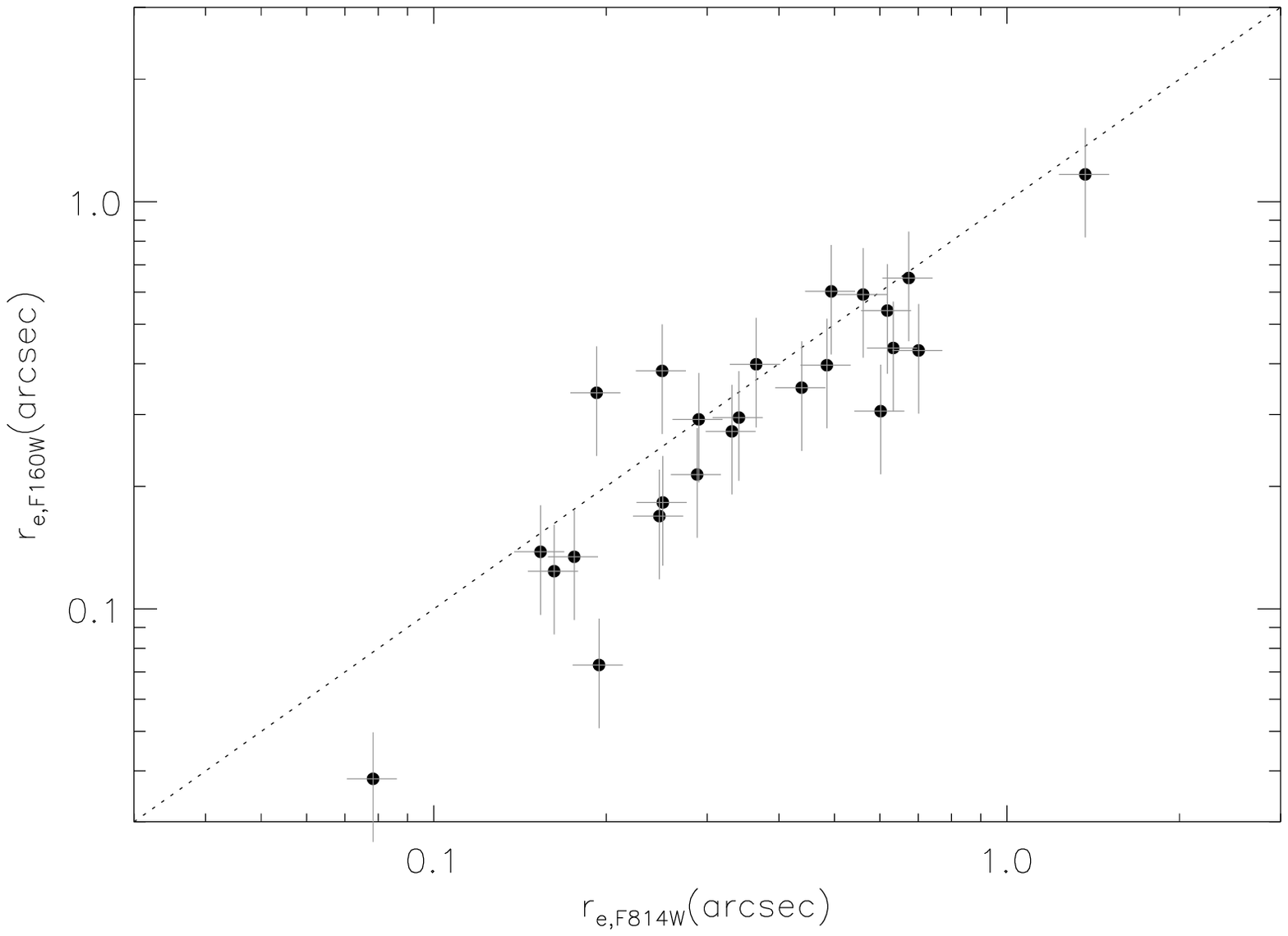,width=0.45\textwidth} 

\caption{Comparison between the size estimates using the ACS I-band filter vs. the NICMOS H-band
for 27 galaxies in our sample at $0.8 < z < 1.8$ where NICMOS data is available.}

\label{nicmos}
\end{figure*}

We found that sizes measured in the NICMOS images, and those measured with the ACS I-band data,
are well correlated (see Figure \ref{nicmos}). The scatter between both measurements is 32\%
(1$\sigma$). There seems to be a systematic (although statistically non-significant) bias of
19$\pm$7\% between both measurements: sizes measured in the redder band (H-band) are slightly 
smaller
than those measured in the bluer band (I-band). Consequently, if any K-correction effect is
affecting our results, the size evolution found here using the I-band filter would be an upper
limit. 

Interestingly, this trend of smaller sizes at redder bands
dr$_{\rm e}(\lambda$)/dlog$\lambda$=-0.6($\pm$0.2) is in qualitative agreement to what is found in
nearby galaxies -0.18$>$dr$_{\rm e}$($\lambda$)/dlog$\lambda$$>$-0.25 
(Barden et al. 2005; McIntosh et
al. 2005).  However, due to the large uncertainty on the bias (and consequently in the size
corrections), and because we do not know how this correction could evolve with redshift we
avoid making any K--correction in our results. In any case, it is important to note that the
trend towards smaller sizes observed using the H-band does not strongly affect the main result
of this paper. In fact, the evolution found in this paper would be even stronger using the
near-infrared band sizes at high redshift.

Another source of concern in the size determination is the presence of an Active Galactic Nuclei
(AGN) in the center of the galaxy which can bias our measurement towards smaller sizes (Daddi et
al. 2005). Deep (200 ks per pointing) X-ray observations (L$_{2-10keV}$$\geq$10$^{42}$ erg
s$^{-1}$ at z$\sim$1) from the Chandra telescope in this field (Nandra et al. 2007; Conselice
et al. 2007b) only detect
emission in 35 objects (i.e. $\sim$4.2\% our sample). We remove these potential AGNs from our
sample in what follows, but our essential results are unchanged. 

\section{Selection effects}

In practice, any galaxy survey has a surface brightness limit beyond which the sample is
incomplete. Characterising this limit is particularly important for high-z samples, where the
effects of the cosmological surface brightness dimming are severe. At a given total flux limit,
the surface brightness limit translates into an upper limit on the size in which a galaxy can be
detected. We have explored whether our K$_s$-band selected sample could be incomplete at large
sizes by examining the detectability of our galaxies as a function of their apparent magnitudes
and sizes.

\begin{figure*}
\epsfig{file=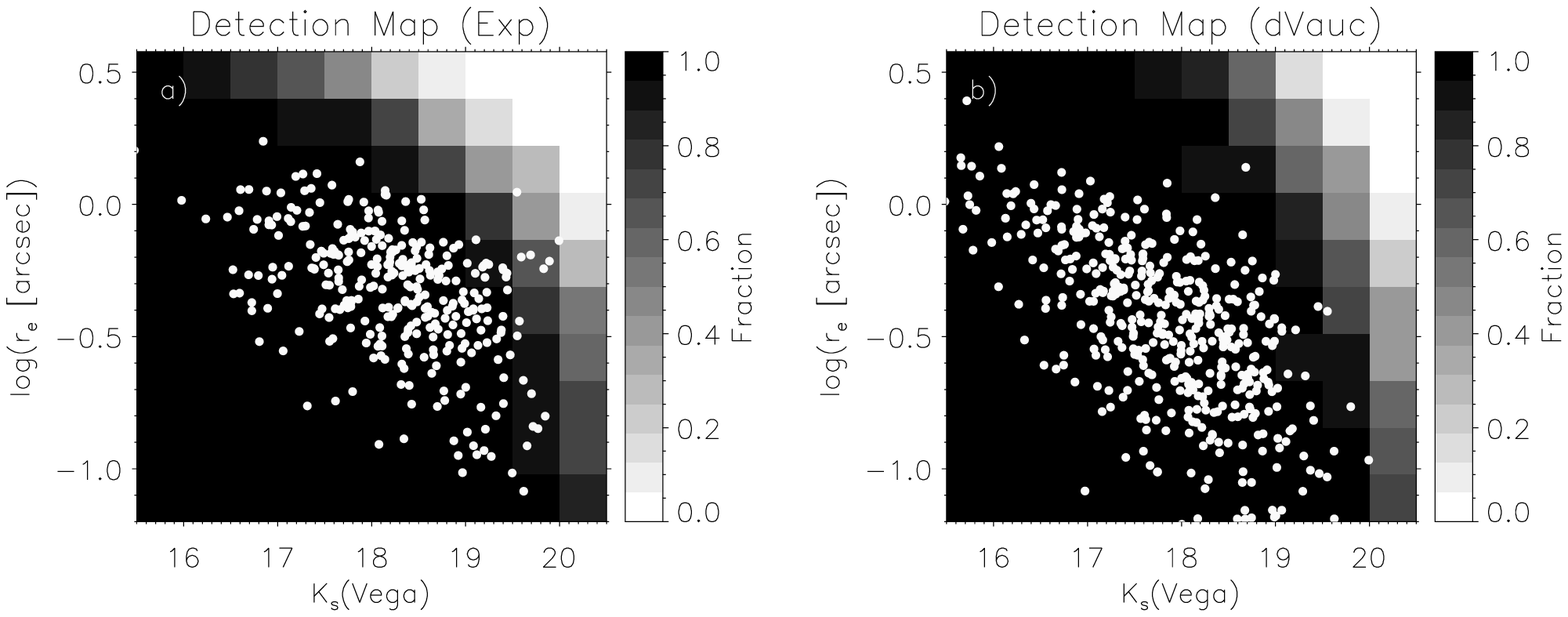,width=0.9\textwidth} 

\caption{(a) Detection map for simulated sources with exponential (n=1) profiles placed at random
in our K$_s$ band image of the Palomar field. The gray-scale map reflects the ratio between input
and recovered objects per input magnitude and log r$_e$ bin. Over-plotted on the map is the
distribution of the sample of K$_s$ band-selected objects in the Palomar field with n$<$2.5 (as
measured in the I-band). Sizes of real objects are those derived in the I-band image. (b) Same as
a) but for simulated sources with de Vaucouleurs (n=4) profiles placed at random in our K$_s$ band
image of the Palomar field.   Over-plotted on the map is the distribution of the sample of K$_s$
band-selected objects in the Palomar field with n$>$2.5 (as measured in the I-band). }

\label{maps}
\end{figure*}

To determine the detection map of the Palomar K$_s$ band image, we have created two sets of 10$^4$
mock sources each with intrinsic exponential or de Vaucouleurs profiles uniformly distributed as
follows: K$_s$ band total magnitudes between 15.5 and 20.5 (Vega) mag (see Fig.
\ref{maghist}), effective radius r$_e$
between 0.0625 and 3.75 arcsec (this will be equivalently to 0.5-30 h$_{70}^{-1}$ kpc at
z$\sim$1), and ellipticities between 0 and 0.8. The simulated sources are placed randomly on the
real image and extracted as for the real source detection. We construct from these simulations
detection maps giving the number of recovered sources over the number of input artificial sources
per input magnitude and input log r$_e$ bin (see Figure \ref{maps}). At it is expected, galaxies
with a de Vaucouleurs profile, and consequently more centrally concentrated, are easier to detect
at a given magnitude.

Over-plotted on the detection maps are the distribution of our sample galaxies. K$_s$ band
magnitudes are measured using the MAG$_{AUTO}$ output from SExtractor. The sizes are those
estimated in the I-band. We are making the implicit assumption that sizes are similar in both
bands (i.e. that K-correction effects are not relevant, see previous Section). At a given
magnitude, the observed size distribution declines more rapidly to larger sizes than the
detection limit. This indicates that our sample is not  affected by incompleteness
for the largest galaxies at a given magnitude.

\section{The observed stellar mass vs size relation}

\begin{figure*}
\centering
\epsfig{file=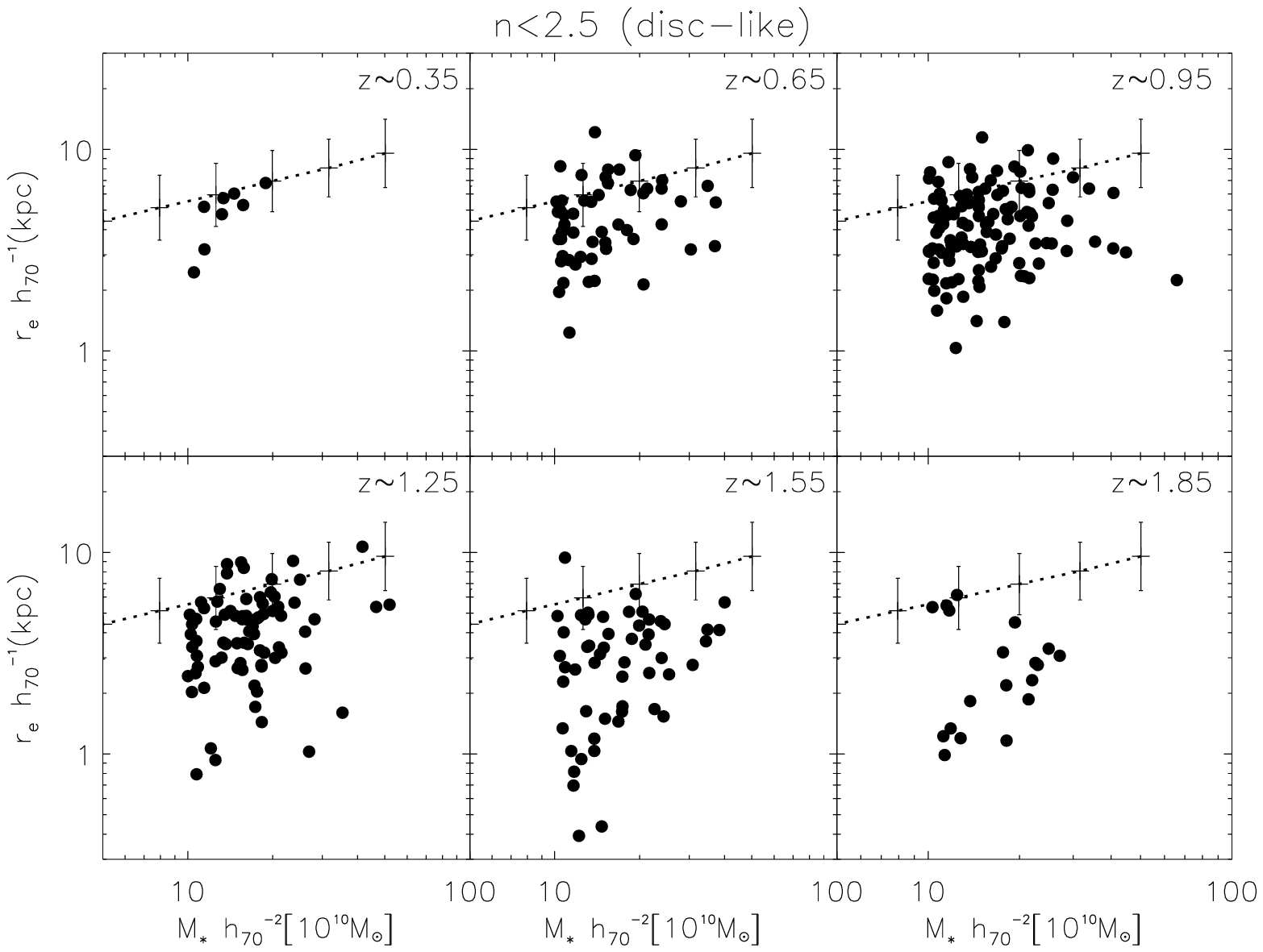,width=0.9\textwidth}

 \caption{Stellar mass-size distribution for our less concentrated (disk-like) galaxies.   
Over-plotted on the observed distribution of points are the mean and dispersion of the distribution
of the S\'ersic half-light radius of the Sloan Digital Sky Survey (SDSS) late-type (n$<$2.5)
galaxies as a function of the stellar mass. We use the SDSS sample as the local reference
(z$\sim$0.1). SDSS sizes were determined also using a circularised S\'ersic model and masses were
retrieved using a Kroupa IMF. SDSS sizes were measured using the observed r'-band, closely
matching the V-band rest-frame filter at z$\sim$0.1. For clarity, individual error bars are not
shown. The mean size relative error is $<$11\%. Uncertainties in the stellar masses are $\sim$0.2
dex.}

\label{disks}
\end{figure*}

\begin{figure*}
\centering
\epsfig{file=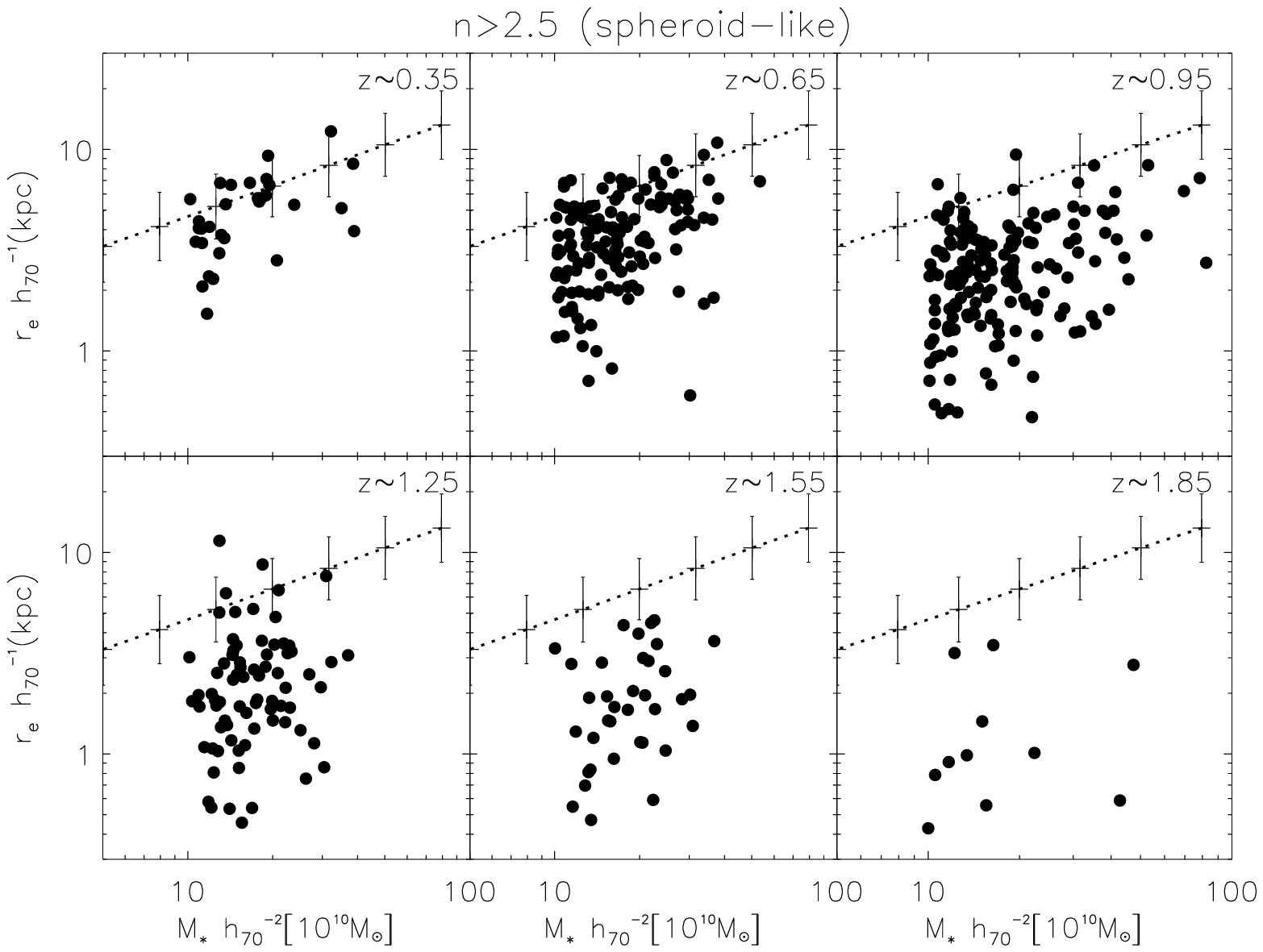,width=0.9\textwidth}

 \caption{Stellar mass-size distribution of our high concentrated (spheroid-like) galaxies.
Over-plotted on the observed distribution of points are the mean and dispersion of the distribution
of the S\'ersic half-light radius of the SDSS early-type (n$>$2.5) galaxies as a function of the
stellar mass. For clarity, individual error bars are not shown. The mean size relative error is
$<$30\%. Uncertainties in the stellar mass are $\sim$0.2 dex.}

\label{spheroids}
\end{figure*}

The accuracy of the structural parameters and the completeness of our sample has been
demonstrated in the previous section. In this section we discuss the stellar mass--size 
relations for our massive galaxy sample. These relations are presented in Figures \ref{disks} (n$<$2.5; 
disk-like objects) and
\ref{spheroids} (n$>$2.5; spheroid-like objects). In each of these figures, we have divided our
sample (over 0.2$<$z$<$2) into six redshift slices with a width of dz=0.3. 

Over-plotted on our observed distributions are the mean and dispersion of the distribution of
the S\'ersic half-light radii from the SDSS (York et al. 2000) galaxies. We use the  SDSS
sample as the local reference. As it has been done for our high-z galaxies, the SDSS galaxy'
sizes were determined from a S\'ersic model fit (Blanton et al. 2003). The characteristics of
the SDSS sample used here are detailed in Shen et al. (2003). The mean of the SDSS galaxies'
redshift distribution used for comparison is 0.1. We use the sizes and the shapes estimated in
the observed SDSS r-band as this closely matches the V-band rest-frame filter at z$\sim$0.1.
SDSS stellar masses were derived using a Kroupa (2001) IMF, which gives the same
masses as the Chabrier IMF used to measure the higher redshift sample stellar masses.

Figures \ref{disks} and \ref{spheroids} show that at a given stellar mass galaxies are
progressively smaller at higher redshift. This evolution is particularly strong for more 
concentrated
(n$>$2.5) galaxies.   We find the remarkable result that there are no spheroid-like objects at
z$>$1.5 on the local relation.

\begin{figure*}
\epsfig{file=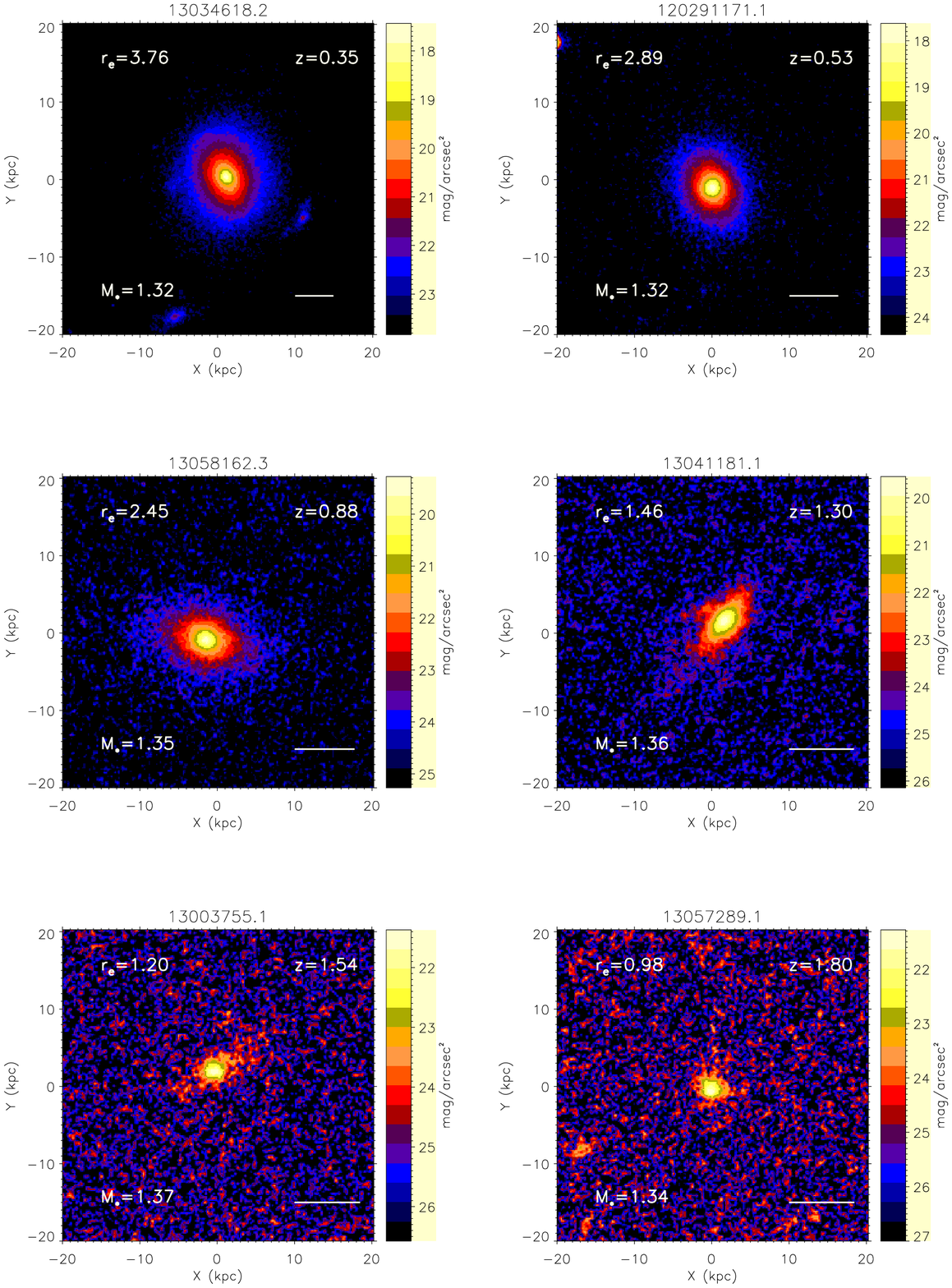,width=0.9\textwidth} 

\caption{Visual structural evolution of equal stellar mass galaxies at different look-back 
times. The panel shows
six concentrated (n$>$2.5, spheroid-like) representative galaxies in our sample at different
redshift (or look-back time, from top-right to bottom-left: 3.9, 5.3, 7.4, 9.0, 9.7 and 10.3 Gyr
back respectively). Effective radii are given in kpc and stellar masses in 10$^{11}$ M$_{\sun}$
units. Galaxies are shown with different surface brightness limits to account for the cosmological
surface brightness dimming. The solid line indicates one arcsec angular size.}

\label{mosaico}
\end{figure*}

To illustrate this size decrease at progressively higher z, Figure \ref{mosaico} shows the
appearance of six of our n$>$2.5 (i.e. spheroid-like) galaxies with the same stellar mass, but
at different look-back times.  The objects shown in Figure \ref{mosaico} have the mean
structural properties of the galaxy population at each of the different redshift slices (see
the stellar mass-size relations in the above redshift intervals in Figures \ref{disks} and
\ref{spheroids}). To allow a fair comparison between the sizes of objects at very different
redshifts, the limiting surface brightness in each panel is changed according to the
cosmological surface bright dimming $\sim$(1+z)$^4$. For this reason the surface brightness of
the object at z=0.35 is shown down to $\sim$3.25 mag brighter than the object at z=1.85.  This
figure visually illustrates that the most massive galaxies are progressively smaller at increasing
redshift.

\begin{figure*}
\centering
\epsfig{file=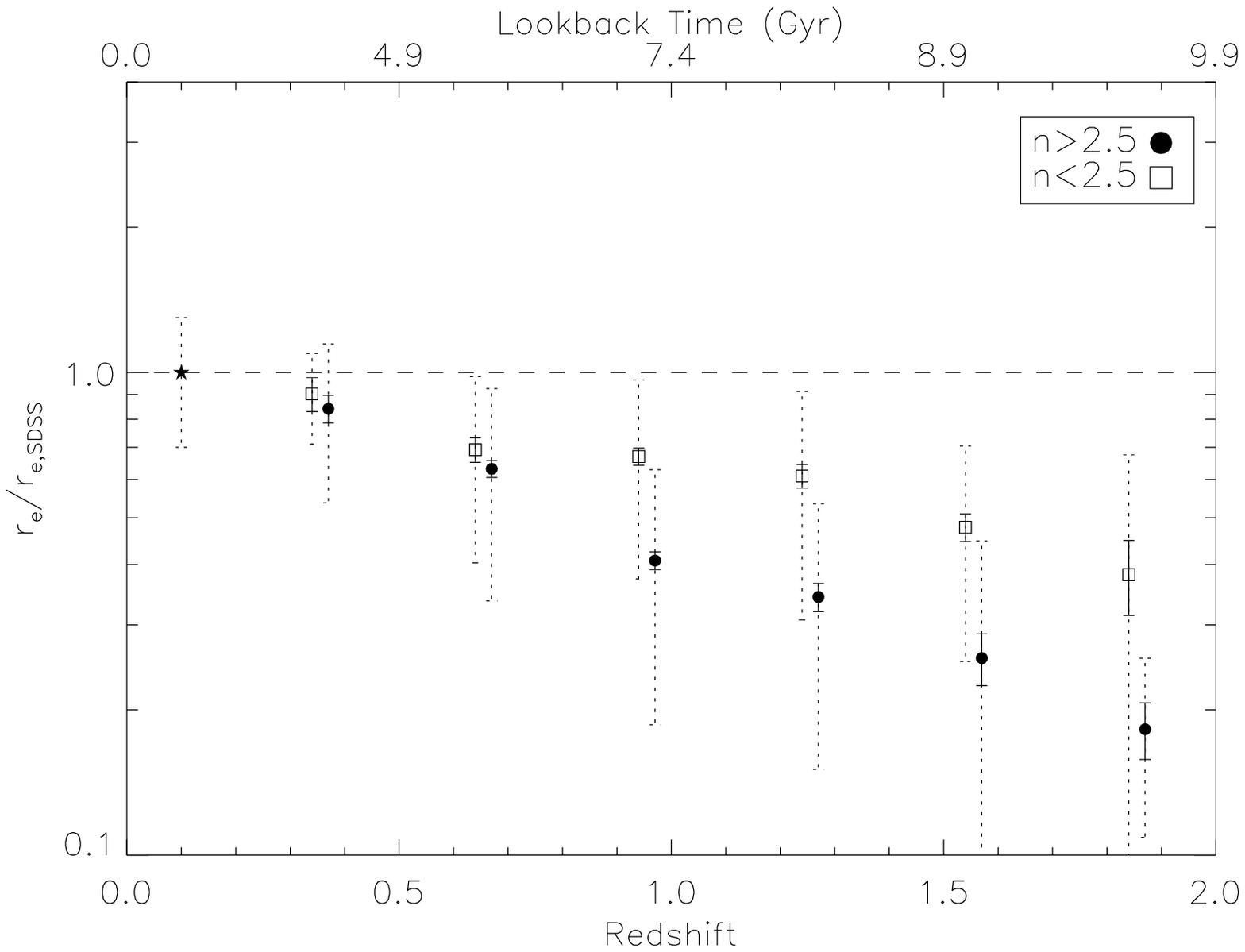,width=0.45\textwidth}

 \caption{Size evolution of the most massive galaxies with lookback time. The evolution with
redshift of the median ratio between the sizes of the galaxies in our sample and the galaxies of
the same stellar mass in the SDSS local comparison sample is shown. Solid points indicate the size
evolution of spheroid-like (n$>$2.5) galaxies. Open squares show the evolution for disk-like
(n$<$2.5) galaxies. The small error bars indicate the uncertainty (1 $\sigma$) at the median
position. The large error bars represent the dispersion in the distributions.}

\label{evolution}
\end{figure*}

To quantify the size evolution, we measure the ratio between the observed size, and the expected
size at a given stellar mass by comparing with the SDSS (Shen et al. 2003) distribution at
different redshifts. To estimate the expected size from SDSS at a given stellar mass, we
interpolate linearly between the SDSS points when necessary. The evolution of the median and the
dispersion of the above ratio are shown in Figure \ref{evolution} and listed in Table \ref{data}. 
The observed size evolution of  M$_{\star}$$>$10$^{11}$ M$_{\sun}$ galaxies is stronger than the
one found in previous work using less massive galaxies (see the detailed comparison of this
issue in the next section). Following recent claims (i.e. Maraston et al. 2006) that stellar masses
could be systematically overstimated by a factor of few at high z, we have repeated our analysis
under the assumption that our masses could be overestimated by a factor of 2. In this case, at
z$\sim$1.5, our galaxies will still be more compact than present-day galaxies of the same stellar
mass by a factor of 1.6 for n$<$2.5 and a factor of 3 for n$>$2.5. In other words, even a
systematic effect of factor of 2 in the stellar mass determination can not avoid a significant
evolution in the size of these galaxies. In fact, to prevent a significant evolution of the sizes
of galaxies at a given stellar mass we found that, at z$\sim$1.5, our stellar masses would need to
be overestimated by a factor of $\sim$10  for galaxies with n$<$2.5 and overestimated by a factor
of $\sim$50 for galaxies with n$>$2.5.

\begin{table*}
 \centering
 \begin{minipage}{140mm}

  \caption{Size evolution of the most massive (M$_{\star}$$>$10$^{11}$ M$_{\sun}$) galaxies in
the Universe}

  \begin{tabular}{ccccc}
  \hline
Redshift Range & n$<$2.5 & n$<$2.5 & n$>$2.5 &  n$>$2.5  \\
               &  $<$r$_e$/r$_{e,SDSS}$$>$($\pm$1$\sigma$)  & Dispersion 
					 &  $<$r$_e$/r$_{e,SDSS}$$>$($\pm$1$\sigma$) &  Dispersion \\
 \hline
0.1 (SDSS) & 1 		   & 0.30 & 1  		  & 0.30  \\
0.2-0.5	  & 0.90(0.07) & 0.19 & 0.84(0.06) & 0.30  \\
0.5-0.8	  & 0.69(0.04) & 0.29 & 0.63(0.03) & 0.30  \\
0.8-1.1	  & 0.67(0.03) & 0.29 & 0.41(0.02) & 0.22  \\
1.1-1.4	  & 0.61(0.03) & 0.30 & 0.34(0.02) & 0.19  \\
1.4-1.7	  & 0.48(0.03) & 0.23 & 0.26(0.03) & 0.19  \\ 
1.7-2.0	  & 0.38(0.07) & 0.30 & 0.18(0.02) & 0.07  \\
\hline
\label{data}
\end{tabular}
\end{minipage}
\end{table*}

\section{Comparison with other samples}

\begin{figure*}
\centering
\epsfig{file=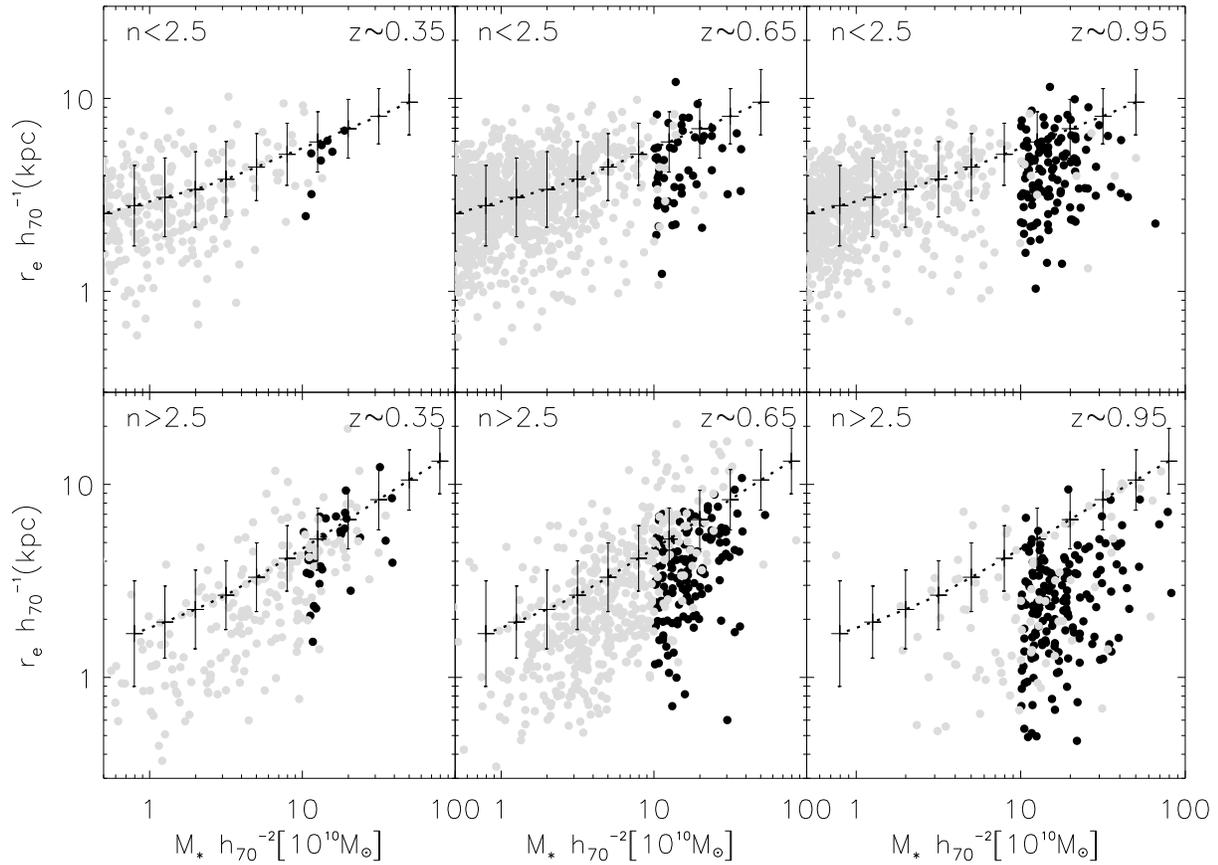,width=0.9\textwidth}

 \caption{Stellar mass-size evolution with redshift of disk-like (n$<$2.5) and spheroid-like
(n$>$2.5) galaxies up to z$\sim$1. The Palomar massive galaxies sample is shown (solid points) in
comparison with the lower stellar masses galaxies from the GEMS survey (gray symbols). Over-plotted
are the stellar mass - size relation from the SDSS (Shen et al. 2003).}

\label{gemspalomar}
\end{figure*}

An interesting point to explore  is how the size evolution found here for the most
massive galaxies compares with the size evolution observed for galaxies with lower masses. This
comparison can be done in detail in the redshift range 0$<$z$<$1 using the data from one of the
largest sample currently available: the GEMS survey (Rix et al. 2004). The stellar mass-size
relation of this survey has been already derived for late-type (Barden et al. 2005) and
early-type (McIntosh et al. 2005) galaxies. The GEMS late- and early-type separation criteria
are based on the S\'ersic index n. Late-types are defined through n$<$2.5, and early types
through n$>$2.5, and a colour within the ``red sequence'' (Bell et al. 2004). The comparison of the
distribution of their data points with ours are shown in Figure \ref{gemspalomar}.

\begin{figure*}
\centering
\epsfig{file=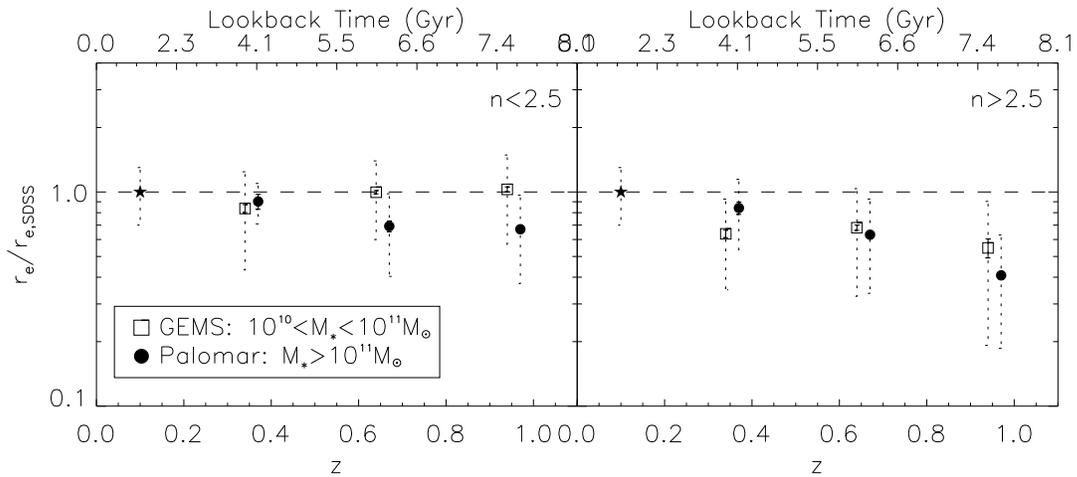,width=0.8\textwidth}

 \caption{Redshift evolution of the ratio between the observed size and the present-day mean size
at a given stellar mass. Open squares represent those galaxies within the mass ranges
10$^{10}$$<$M$_{\star}$$<$10$^{11}$ M$_{\sun}$ (GEMS), whereas solid points are
M$_{\star}$$>$10$^{11}$ M$_{\sun}$ (Palomar) galaxies.}

\label{gemspalomarevol}
\end{figure*}

It is encouraging to see that these two independent analyses and data sets match  well where
there is overlap in their stellar mass (i.e. for the small subset of GEMS galaxies with
M$_{\star}$$>$10$^{11}$ M$_{\sun}$). These two data sets allow us to compare the difference in
size evolution in the 0$<$z$<$1 redshift interval at two different mass ranges
10$^{10}$$<$M$_{\star}$$<$10$^{11}$ M$_{\sun}$ (GEMS), and M$_{\star}$$>$10$^{11}$ M$_{\sun}$
(Palomar). This is shown in Figure \ref{gemspalomarevol}. From this comparison we can see that
more massive galaxies evolve in size much faster than lower mass objects (particularly for
disk-like galaxies). This mass dependent evolution was hinted in previous works (Trujillo et
al. 2006a) but our current large data set show this more clearly and robustly.

At higher redshift, the amount of data is more limited. At 1$<$z$<$2 our results are in good
agreement with recent findings (based on a few objects) of massive compact galaxies at high-z
(Waddington et al. 2002; Daddi et al. 2005; di Serego Alighieri et al. 2005; Trujillo et al. 2006b;
Longhetti et al. 2007). For example, using the 10  of the most massive galaxies (1.2$<$z$<$1.7) in
the MUNICS survey, Trujillo et al. (2006b) found that these galaxies were a factor of
4$_{-1.0}^{+1.9}$ smaller that local counterparts. At even higher z (i.e. z$\sim$2.5) there has
been also recent claims of very compacts (r$_e$$\lesssim$1 kpc) massive galaxies (Trujillo et al.
2006a; Zirm et al. 2007; Toft et al. 2007).

\section{Discussion}

As shown in  Section 4 the compact nature of our most massive galaxies at high-z can not be
interpreted as a K-correction or AGN effect. In addition, there is no observational evidence
in the local Universe for galaxies as massive and compact as the ones in our sample. These two
observational facts  raise the following two questions: first, how can these objects be so
dense in the past? Second, how do these objects evolve in stellar mass and/or size in
order to reach the current local relation?

Addressing the first question, recent theoretical results suggest that major galaxy mergers in
the early Universe had a much larger component of cold gas available than in the present
(Khochfar \& Silk 2006a). These wet (dissipative) mergers generate very efficient and massive
starbursts creating a very compact massive remnant.  Consequently, the very dense nature of our
objects at high-z could reflect the much denser condition of the Universe at the time
of their formation.

Focusing on the second question, at lower redshifts the available amount of gas is less, and new
`dry' (dissipationless) mergers (van Dokkum 2005; Bell et al. 2006) would be the dominant
mechanisms of size and stellar mass growing (Ciotti \& van Albada 2001; Nipoti, Londrillo  \&
Ciotti 2003; Khochfar \& Burkert 2003; Dom\'inguez-Tenreiro et al. 2006; Naab, Khochfar \& Burkert
2006; Ciotti, Lanzoni \& Volonteri 2007). Dry mergers are not efficient at forming new stars, but
are efficient in increasing the size of the objects. A particular effective size evolutionary
mechanism (r$_e$$\sim$M$_{\star}^{1.3}$) has been recently found (Boylan-Kolchin, Ma \& Quataert
2006) in mergers of galaxies with radial orbits along large-scale structure filaments. This
kind of mechanism would be able to evolve our compact galaxies (a factor of 4 smaller at
z$\sim$1.5) to the local relation with just two major (equal-mass) mergers. Since $z \sim 2$
very few gas rich mergers occur in massive galaxies (e.g., Conselice et al. 2003; Conselice 2006),
but a few dry mergers are possible based on pair counts (e.g., Lin et al. 2004). In addition, dry
mergers of similar mass objects (and consequently, potentially similar ages and metallicities) will
also help to understand the age-uniformity found in the local massive (spheroid) galaxies. On the
other hand, in agreement with our results here, new semi-analytical models (Somerville et al. 2007)
find that disc-like galaxies (i.e. those presumably with a relative quiet live) evolve only mildly
in size since high-z.

Summarising our results, we find that the size evolution of the most massive galaxies is not 
consistent with a scenario whereby massive galaxies were fully assembled in the early 
Universe, and have subsequently evolved passively until today (i.e. a pure "monolithic" scenario). 
In fact, our
findings agree with a scenario where a  fraction of the most massive galaxies
possibly formed in a short ``monolithic-like'' collapse and then evolved through major
gas-rich, or gas-poor merging. In this sense, the two scenarios would be just different phases
of galaxy formation and evolution.  Our results, consequently, point to a scenario whereby the
stellar populations of the most massive galaxies we observe today were located in different
``primordial'' massive galaxy pieces in the early Universe.

\section{Acknowledgements}

Authors are grateful to M. Barden and D. McIntosh for providing us with the GEMS data points to
allow us a comparison of our results with their less massive galaxies. We acknowledge useful
discussions with O. Almaini and A. Arag\'on-Salamanca. We also thank the useful suggestions from
an anonymous referee. The Palomar and DEEP2 surveys would not have been completed without the
active help of the staff at the Palomar and Keck observatories. The Palomar Survey was supported by
NSF grant AST-0307859 and NASA STScI grant HST- AR-09920.01.A. Support for the ACS imaging of the
EGS in GO program 10134 was provided by NASA through NASA grant HST-GO-10134.13-A from the Space
Telescope Science Institute.

\appendix

\onecolumn

\section{Structural parameters of the Palomar/DEEP2 sample massive galaxies}

NOTE: FULL DATA TABLE IS AVAILABLE ON THE ONLINE VERSION OF MNRAS OR BY REQUEST TO THE AUTHORS.

The following table contains the information about the structural parameters of the sample of
galaxies analysed in this paper. Column 1 indicates the galaxy identification, column 2 lists 
the apparent K-band magnitude in the Vega system, column 3 is the efffective radius along the
semimajor axis, column 4 indicates the value of the S\'ersic index of the fit, column 5 is the
ellipticity of the source, column 6 is the stellar mass of the galaxy in units of
10$^{10}$h$_{70}$$^{-2}$M$_{\sun}$, column 7 is the measured redshift of the object, column 8 and 9
are the R.A. and Dec. of the source and, finally, column 10 specifies whether the redshift was
determined spectroscopically (1) or photometrically (0).

\begin{longtable}{cccccccccc}
  \caption{Properties of the Palomar/DEEP2 sample galaxies} \label{rawdata} \\
  \hline
 \multicolumn{1}{c}{Galaxy ID}&\multicolumn{1}{c}{K$_s$}&  
 \multicolumn{1}{c}{a$_e$}&\multicolumn{1}{c}{n}& 
 \multicolumn{1}{c}{$\epsilon$}&\multicolumn{1}{c}{M$_{\star}$}& 
  \multicolumn{1}{c}{z}&\multicolumn{1}{c}{R.A.}&\multicolumn{1}{c}{Dec.}&\multicolumn{1}{c}{Spec.}\\
  \multicolumn{1}{c}{}&\multicolumn{1}{c}{(Vega mag)}&  \multicolumn{1}{c}{(arcsec)}&
  \multicolumn{1}{c}{}&  \multicolumn{1}{c}{}&
  \multicolumn{1}{c}{(10$^{10}$h$_{70}$$^{-2}$M$_{\sun}$)}&  
  \multicolumn{1}{c}{}&\multicolumn{1}{c}{(J2000)}&\multicolumn{1}{c}{(J2000)}&\multicolumn{1}{c}{}\\
 \endfirsthead

\multicolumn{10}{c}{{ \tablename\ \thetable{} -- continued from previous page}} \\
 \hline
 \multicolumn{1}{c}{Galaxy}&\multicolumn{1}{c}{K$_s$}&  
 \multicolumn{1}{c}{a$_e$}&\multicolumn{1}{c}{n}& 
 \multicolumn{1}{c}{$\epsilon$}&\multicolumn{1}{c}{M$_{\star}$}& 
  \multicolumn{1}{c}{z}&\multicolumn{1}{c}{R.A.}&\multicolumn{1}{c}{Dec.}&\multicolumn{1}{c}{Spec}\\
  \multicolumn{1}{c}{}&\multicolumn{1}{c}{(mag)}&  \multicolumn{1}{c}{(arcsec)}&
  \multicolumn{1}{c}{}&  \multicolumn{1}{c}{}&
  \multicolumn{1}{c}{(10$^{10}$h$_{70}$$^{-2}$M$_{\sun}$)}&  
  \multicolumn{1}{c}{}&\multicolumn{1}{c}{(J2000)}&\multicolumn{1}{c}{(J2000)}&\multicolumn{1}{c}{}\\
  \hline
  \endhead
  
  \hline \multicolumn{10}{c}{{Continued on next page}} \\ \hline
\endfoot

\hline \hline
\endlastfoot

 \hline
   11051619 & 16.75 & 0.90 & 3.60 & 0.38 & 11.94 & 0.46 & 214.258865 & 52.399944  &    0	 \\
   12004106 & 17.13 & 0.80 & 6.72 & 0.11 & 26.72 & 0.74 & 214.379730 & 52.443295  &    1	 \\
   12004308 & 16.95 & 0.89 & 1.23 & 0.58 & 18.03 & 0.64 & 214.266815 & 52.412926  &    0	 \\
   12004443 & 17.06 & 0.48 & 5.04 & 0.21 & 26.96 & 0.77 & 214.318298 & 52.429092  &    1	 \\
\hline
\end{longtable}

\twocolumn

\end{document}